# The Non-Uniform Distribution of Galaxies from Data of the SDSS DR7 Survey

**A. O. Verevkin, Yu. L. Bukhmastova, and Yu. V. Baryshev**

*Sobolev Astronomical Institute, St. Petersburg University, Russia*
*(yubaryshev@mail.ru , bukh_julia@mail.ru)*

**Abstract**—We have analyzed the spatial distribution of galaxies from the release of the Sloan Digital Sky Survey of galactic redshifts (SDSS DR7), applying the complete correlation function (conditional density), two-point conditional density (cylinder), and radial density methods. Our analysis demonstrates that the conditional density has a power-law form for scales lengths 0.5–30 Mpc/h, with the power-law corresponding to the fractal dimension $D = 2.2 \pm 0.2$; for scale lengths in excess of 30 Mpc/h, it enters an essentially flat regime, as is expected for a uniform distribution of galaxies. However, in the analysis applying the cylinder method, the power-law character with $D = 2.0 \pm 0.3$ persists to scale lengths of 70 Mpc/h. The radial density method reveals inhomogeneities in the spatial distribution of galaxies on scales of 200 Mpc/h with a density contrast of two, confirming that translation invariance is violated in the distribution of galaxies to 300 Mpc/h, with the sampling depth of the SDSS galaxies being 600 Mpc/h.

## 1. INTRODUCTION

The current spatial distribution of galaxies contains important information on the origin and evolution of the large-scale structure (LSS) of the Universe; it is a crucial test for modern cosmological models [1]. During almost the entire 20th century, only two-dimensional (2D) distributions of galaxies on the celestial sphere were available for studies of the LSS. These were analyzed using angular two-point correlation functions [2, 3]. At that time, a power-law form was established for the spatial correlation function, $\xi(r) = (r/r_0)^{-\gamma}$, with the characteristic linear scale length for inhomogeneity $r_0 = 5$ Mpc/h and the characteristic slope $\gamma = 1.77$. In these studies, the distribution of galaxies in the Universe was assumed to be homogeneous beginning with the scale length $3r_0 = 15$ Mpc/h [4].

Only toward the end of the 1980s did large numbers of galaxy redshifts become available, and the first three-dimensional (3D) galaxy distributions were plotted. The transition from analyses of 2D distributions to the real space distribution of galaxies revealed many unexpected results [3]. Two problems came to the fore: the scale for inhomogeneity and differences of the power-law exponent from values derived from 2D catalogs of galaxies. Thus, correlation analyses of redshift surveys completed by the late 1990s, such as that presented in [5], demonstrated that the complete correlation function had a power-law form. The exponent was $\gamma = 1.0 \pm 0.2$ to scales of about 100 Mpc/h, while the scale length for inhomogeneity of the spatial distribution of the galaxies can considerably exceed this value.

In recent years, two deep wide-angle surveys of galaxy redshifts were completed: the 2dF survey (http://www2.aao.gov.au/2dFGRS/) and the Sloan Digital Sky Survey (SDSS) (http://www.sdss.org/). These contain about $10^6$ galaxies (compared to the $10^3 - 10^4$ galaxies in the surveys of the 1980s).

The current paper presents our analysis of the spatial distribution of galaxies based on the SDSS DR7 data. We apply correlation methods as well as radial distribution methods, which can be used to identify the largest-scale inhomogeneities, on scale lengths comparable to the depth of the considered sample. The analysis methods are presented in Section 2. Section 3 describes the galaxy sample from the SDSSDR7. Section 4 is devoted to our application of the conditional density, two-point conditional density, and radial density methods. Section 5 discusses our results and the main conclusions.

## 2. ANALYSIS METHODS APPLIED TO THE SPATIAL DISTRIBUTION OF GALAXIES

Various statistical methods can be applied to the analysis of 2D and 3D catalogs of galaxies. Comprehensive reviews of the mathematical methods used to describe the large-scale distribution of galaxies can be found in the monographs [2, 6, 7]. It is very important to use mathematical techniques that are adequate to the real structures when analyzing the distribution of galaxies.

The two-point correlation function is a widely used classical approach for the analysis of the LSS. This method was taken from the statistical physics of density fluctuations of ordinary gas, further developed, and widely applied to galaxy data [2, 8]. The correlation function is a measure of the deviation of the observed number of galaxy pairs from a uniform distribution expressed with a Poisson law. The various methods used to estimate the two-point correlation function for a distribution of points differ in their approach to taking into account the sample's boundary conditions [9]. An important assumption of the method is that homogeneity is already present within the galaxy sample and that it corresponds to the mean universal density of galaxies.

The two-point correlation function works well when studying fluctuations in a "smooth" sample on small scale lengths. However, actual observations possess large-scale inhomogeneities, with the density of the sample objects varying with distance along the line of sight and/or direction on the celestial sphere.

An alternative that does not assume a priori homogeneity of the distribution is a fractal approach to studies of the LSS of the Universe [3, 7]. During the last three decades, the mathematical concept of fractality has appreciably influenced many scientific disciplines. If a fractal structure is confirmed for very large scale lengths, additional studies become necessary in order to develop a self-consistent view of the evolution of the Universe that created the observed
structures. The conditional density method for spherical shells and the integrated conditional density method for spheres, also taken from statistical physics, can successfully be applied in a fractal approach. The advantage of these methods is that they provide an unbiased estimate of the intrinsic degree of correlation and fractal dimension in the case of large density fluctuations. They can also be used to find an unbiased homogeneity scale length for the galaxy sample, but only when there are no inhomogeneities comparable to the sample depth.

The presence of a flat section in the conditional density is both a necessary and a sufficient condition for the detection of homogeneity.

A strong limitation for the practical application of the conditional density method is the requirement that the volume of the analyzed sample contain enough space for a complete sphere. For example, galaxy surveys with a thin-slice geometry cannot be used to measure the conditional density on scale lengths exceeding the thickness of the slice; i.e. the diameter of the largest sphere completely embedded in the survey volume. The methods that are usually applied to such scale lengths are those that are most effective for the geometric conditions of actual galaxy surveys, such as the two-point conditional radial density method [10].

Stochastic and systematic errors due to the limited volumes of the samples considered are inherent to the methods used in modern studies. There currently exist analytical methods for estimating the errors of the two-point correlation function and conditional density in the case of uniform distributions. No analytical error-estimation techniques have been developed for strongly irregular (also fractal) structures, and the only possibility for analyzing these errors are numerical experiments using model distributions of points with known properties.

## 2.1. Definitions of Correlation Functions

The theory of stochastic processes introduces and analyzes various functions aimed at correlation analyses [3, 7].

The complete two-point correlation function $R_{\mu\mu}$ (or simply the complete correlation function) for a stationary isotropic process $\mu(r)$ is defined as

$$R_{\mu\mu}(r) = \langle \mu(\vec{r}_1)\mu(\vec{r}_2) \rangle \qquad (1)$$

$r = |\vec{r}| = |\vec{r}_1 - \vec{r}_2|$ -is the distance between the considered points and angular brackets, $<>$, denote the mathematical expectation value for all realizations of a stochastic process.

Taking into account the intrinsic constant mean value of the process $\mu_0$,

$$\mu_0 = \langle \mu(\vec{r}) \rangle = const \qquad (2)$$

we can define the two-point correlation function $C_2$ for fluctuations about $\mu_0$ (or simply, the reduced correlation function) as

$$C_2(r) = \langle (\mu(\vec{r}_1) - \mu_0)(\mu(\vec{r}_2) - \mu_0) \rangle = R_{\mu\mu}(r) - \mu_0^2 \qquad (3)$$

For $r = 0$, this gives the dispersion of the process $\sigma_\mu^2 = C_2(0)$. The difference between the complete and reduced correlation functions, $R_{\mu\mu}$ and $C_2$, is important. For a stochastic process with power-law, long-distance correlations, the complete correlation function $R_{\mu\mu}$ has a power-law form while the reduced correlation function $C_2$ cannot be a power law in this case, according to (3).

## 2.2. The Conditional Density Method

In the case of a continuous stochastic process, the conditional density $\Gamma(r)$ can be expressed via the complete correlation function (1) as

$$\Gamma(r) = \frac{R_{\mu\mu}(r)}{\mu_0} = \frac{\langle \rho(\vec{r}_1)\rho(\vec{r}_1 + \vec{r}) \rangle}{\rho_0} \qquad (4)$$

Here, $\rho(r)$ is the stochastic density field and $\rho_0$ is the density averaged over the ensemble. The $\Gamma$ function has the physical dimension of density (g/cm3) and is a measure of correlation in the total-density field, without subtraction of the mean density. The physical dimension of the $\Gamma$ function agrees with the general interpretation of $\Gamma(r)$ as the mean-density law about each point of the structure. Thus, this estimate becomes a natural detector of fractal structure.

For all the analyzed quantities, we must distinguish between their analytical definitions for the case of an infinite unlimited process (ensemble) and estimated functions applied to real finite distributions (samples) taking into account the boundary geometry.

Consider a discrete stochastic process whose realizations represent sets of particles located at random position $\{\vec{r}_i\}, i = 1, \ldots, N$, so that the realized particle-number density $n(r)$ is given by

$$n(r) = \sum_{i=1}^{N} \delta(\vec{r} - \vec{r}_i) \qquad (5)$$

The statistical quantities calculated for different realizations of the finite sample of an ensemble display fluctuations. Below, we consider ergodic, random processes; i.e., processes for which averaging over different realizations is equivalent to averaging over an infinite volume (set of points) for a single realization of the ensemble.

**2.2.1. Integrated conditional density.**

The integrated conditional density $\Gamma*(r)$ is defined as

$$\Gamma*(r) = \langle n(r' < r) \rangle_P = \frac{\langle N(r) \rangle_P}{\|C(r)\|} \tag{6}$$

$\langle n(r' < r) \rangle$ - is the mean density in a sphere of radius $r$ about an arbitrary origin; $N(r) = \int_0^r n(r') d^3 r'$ – is the number of points within the sphere; $\|C(r)\| = \frac{4}{3} \pi r^3$ – is the volume of the sphere; and $\langle ... \rangle_P$ denotes averaging over all points of the ensemble, provided that the centers of the spheres are placed at points of the ensemble. The integrated conditional density is estimated as [7]:

$$\Gamma_E*(r) = \frac{\overline{N(r)_P}}{\|C(r)\|} = \frac{1}{N_c(r)} \sum_{i=1}^{N_c(r)} \frac{N_i(r)}{\|C(r)\|} \tag{7}$$

where $N_c(r)$ is the number of points (centers) considered with shells that are completely contained within the sample volume, and $\langle ... \rangle_P$ denotes averaging over all points of the sample. In our case, we take into account the boundary based on the method of complete spherical shells. This means that the averaging uses only points where a sphere of a given radius can be drawn completely within the studied sample volume.

Conditional density estimates for the simplest uniform process yield a constant corresponding to the mean density of the ensemble:

$$\Gamma*(r) = \langle n \rangle = const \tag{8}$$

The uncertainty in the integrated conditional density, $\delta \Gamma*(r) = \Gamma_E*(r) - \Gamma*(r)$ is estimated as

$$\delta \Gamma*(r) \simeq \sqrt{\frac{\overline{n}}{N_c(r) \|C(r)\|}} \sim r^{3/2}, r < n^{-1/3}$$

$$\delta \Gamma*(r) \simeq \frac{1}{V} \sqrt{N} \sim const, r > n^{-1/3} \tag{9}$$

where $N$ is the number of points in the sample and $V$ is the volume of the subsample. Figure 1 shows the curve $\Gamma*(r)$ for a uniform distribution with $N = 4000$ points. The dashed line is the mean density in the sample, $n$. The uncertainties were calculated according to (9).

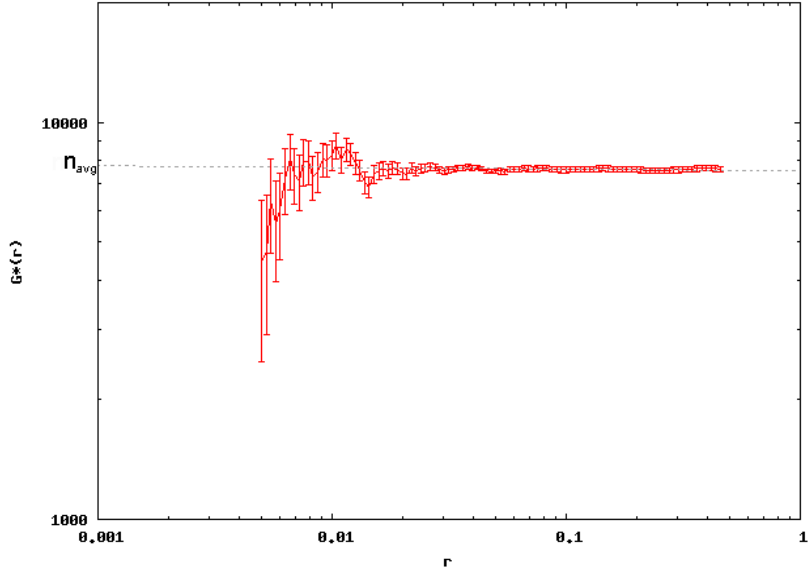

**Fig. 1.** The Γ *(r) relation (Poisson distribution; N = 4000).

### 2.2.2. Conditional density in shells.

The conditional density in shells, Γ(r), is defined as

$$\Gamma(r)=\langle n(r)\rangle_P \simeq \frac{\langle N(r,\Delta r)\rangle_P}{\|C(r,\Delta r)\|} \qquad (10)$$

where $n(r)$ is the mean density in a sphere of radius $r$ around an arbitrary origin, $N(r,\Delta r)=\int_{r}^{r+\Delta r} n(r')d^3r'$ – the number of points in a spherical shell with radius $r$ and thickness $\Delta r$, a $\|C(r,\Delta r)\|=\frac{4}{3}\pi[(r+\Delta r)^3-r^3]$ -the volume of this shell, $\langle...\rangle_P$ denotes averaging over all points of the shell, under the condition that the centers of the spherical shells are located at points of the ensemble (thus the term "conditional"). The conditional density is estimated as [7]:

$$\Gamma_E(r)=\frac{\overline{N(r,\Delta r)_P}}{\|C(r,\Delta r)\|}=\frac{1}{N_c(r+\Delta r)}\sum_{i-1}^{N_c(r+\Delta r)}\frac{N_i(r,\Delta r)}{\|C(r,\Delta r)\|} \qquad (11)$$

$N_c(r+\Delta r)$ - is the number of considered points (centers) with shells completely within the sample volume. The boundary conditions are taken into account as in [7]. As a rule, a logarithmic sequence of shells is used in computations: $\Delta r=ar$, where $a\ll 1$.

The Γ function (10) and its estimate (11) have a power-law form for a fractal structure:

$$\hat{\Gamma}(r)=\Gamma_0 r^{-\gamma} \qquad (12)$$

where

$$\gamma=D-3 \qquad (13)$$

Here, $D$ is the fractal dimension. This property of the estimate of Γ(r) makes it possible to obtain an unbiased estimate of the fractal dimension for a sample of galaxies.

For a uniform distribution,

$$\Gamma(r) = \langle n \rangle = const \qquad (14)$$

The uncertainty in the integrated conditional density $\delta\Gamma(r) = \Gamma_E(r) - \Gamma(r)$ is estimated as

$$\delta\Gamma(r) \simeq \sqrt{\frac{\bar{n}}{N_c(r+\Delta r)\|C(r+\Delta r)\|}} \sim r^{3/2}, r < r_0$$

$$\delta\Gamma(r) \simeq \frac{1}{V}\sqrt{N} \sim const, r > r_0 \qquad (15)$$

where $r_0^2 = \frac{1}{4\pi\bar{n}\Delta r}$ - is the boundary radius beyond which crossings of spherical shells around different points become significant.

Figure 2 displays the Γ(r) curve for a uniform distribution with N = 4000 points. The dashed line is the mean density in the sample, n. The uncertainties were calculated according to (15).

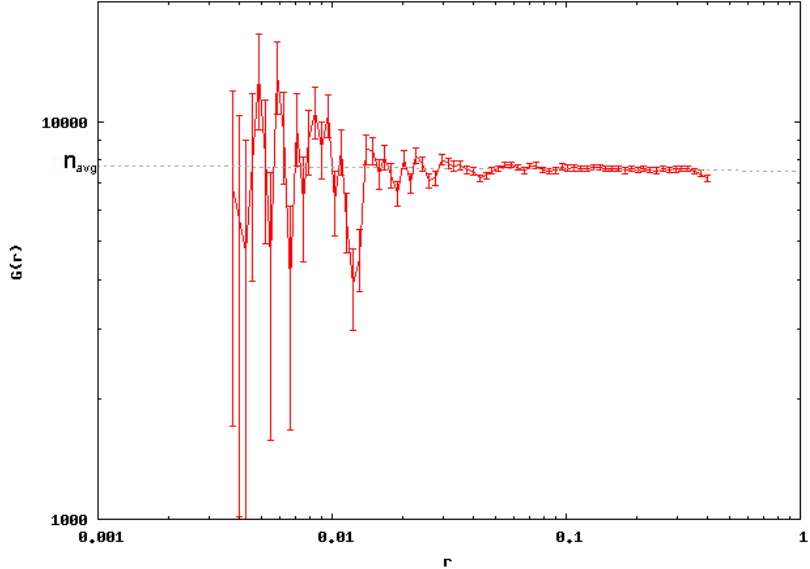

**Fig. 2.** The Γ(r) relation (Poisson distribution; N = 4000).

## 2.3. The Cylinder Method

The above methods for analyzing random processes are single-point methods because the center of the sphere where we count particles is placed at a single point with the coordinates $\{\vec{r}_a\}$. Some cosmological problems require the use of two-point conditional densities, where we fix two particles, $\{a,b\}$ with the coordinates $\{\vec{r}_a, \vec{r}_b\}$. The transition from one point to two-point conditional densities in the analysis of fractal structures is equivalent to the transition from two-point to three-point correlation functions in the analysis of ordinary random processes.

To describe the distribution of particles along a cylinder with its axis passing through two points of the structure $\{a,b\} \subset \{x_i, i=1,\ldots,N\}$, we introduce the two-point conditional radial density of a random process $\eta_{ab}(r)$. According to Mandelbrot's cosmological principle, the particles $a$ and $b$ are statistically equivalent. Thus, the one-point conditional density for each of them is given by

(12), proportional to the probability to find the particles at a distance r from fixed points of the structure.

Let us now take two independent points of the structure separated by a distance $r_{ab}=|\vec{r}_a-\vec{r}_b|$. We introduce the notation C for the event when the particles appear at a distance $r_a$ from the point a and independently at a distance $r_b$ from the point b. Then, C is the combination $A \cup B$ of the two events related to a and b. Thus, the two-point conditional radial density, which is proportional to the probability of detecting particles around the particles a and b, can be expressed as a sum of single-point conditional densities. The first approximation is the assumption that the events are independent. Numerical modeling demonstrated that this approximation was accurate enough for the analysis of ordinary fractal structures [10]. Then, we can express the two-point conditional radial density as [10]

$$n_{ab}=\frac{1}{2}[\Gamma_a(r)+\Gamma_b(r_{ab}-r)]=\frac{DB}{8\pi}r_{ab}^{-\gamma}[(\frac{r}{r_{ab}})^{-\gamma}+(1-\frac{r}{r_{ab}})^{-\gamma}] \qquad (16)$$

where $\gamma=D-3$. The distance r is measured along the straight line joining the particles a and b; at the same time, it determines the radius r of the sphere centered on the first point and the radius $r_{ab}-r$ of the other sphere, centered on the second point. The constant B describes the normalization. In this formula, volume elements are taken along the line joining the two points and, in this sense, r is a one-dimensional Cartesian coordinate.

To estimate $\eta_{ab}(r)$, we can use the statistics

$$n_{ab}(r)=\langle\frac{N_c(x_a,x_b,r,h,\Delta r)}{\pi h^2 \Delta r}\rangle_{\{a,b\}}=\frac{1}{N_{ab}}\sum_{\{a,b\}}^{N_{ab}}\frac{1}{\pi h^2 \Delta r}\int_r^{r+\Delta r}\int_0^h \vec{n}(x) 2\pi h\, dh\, dr \qquad (17)$$

where $N_c$ – is the number of particles in a volume element of a cylinder with diameter h and height $\Delta r$, with its axis joining the structure particles a and b. The volume element is a distance r from a, corresponding to the distance $r_{ab}-r$ from b. Averaging is performed for each pair of points belonging to cylinders with lengths in the interval $(l, l+\Delta l)$.

In practice, a more general situation can be encountered, with the sample simultaneously containing a fractal structure and a uniform background. Then, it is practically possible to calculate the fractal dimension D from the estimate $\eta_{ab}(r)$ by fitting observations with three free parameters, $\gamma, R1, R2$

$$\frac{N(y)}{N}=R1 \cdot \frac{(y^{-\gamma}+(1-y)^{-\gamma})}{2}+R2 \qquad (18)$$

where $N(y)$ – is the observed number of points detected in each "tablet", i.e., within small $(y, y+\Delta y)$ intervals along the cylinder of length l. The variable y is the relative distance measured along the line joining the two points ($y=r/r_{ab}=r/l$). N – the total number of points within cylinders of length l; $\gamma$ the power-law index determining the fractal dimension $D=3-\gamma$, $R1$ –the ratio of the maximum segment length in the sample to the minimum fractal scale length in the sample. The parameter $R2$ takes into account the contribution from a possible Poisson background. We can introduce a measure of the fractal component's relative dominance

$$\beta=\frac{1-R2}{R2} \qquad (19)$$

If $\beta=1$, contributions of the fractal structure and Poisson background are the same ($R2=0.5$).

The main advantage of the cylinder method is that it can be used with surveys in the form of thin slices, and enables analysis of scale lengths comparable to the survey depth. However, for large cylinder lengths, the method begins to feel adverse effects due to the lack of averaging in directions perpendicular to the cylinder lengths, since a long cylinder in a narrow cone can be rotated by only a small angle.

For a uniform distribution, the two-point radial density, like the analogous one-point density, gives the mean density for the ensemble

$$\eta_{ab}(r) = \langle n \rangle = const \tag{20}$$

and thus we obtain the probability density from the normalization condition

$$n(x) = 1 \tag{21}$$

The $n(x)$ relation for a uniform distribution of $N = 4000$ points is presented in Fig. 3, where the theoretical value $n(x) = 1$ is marked.

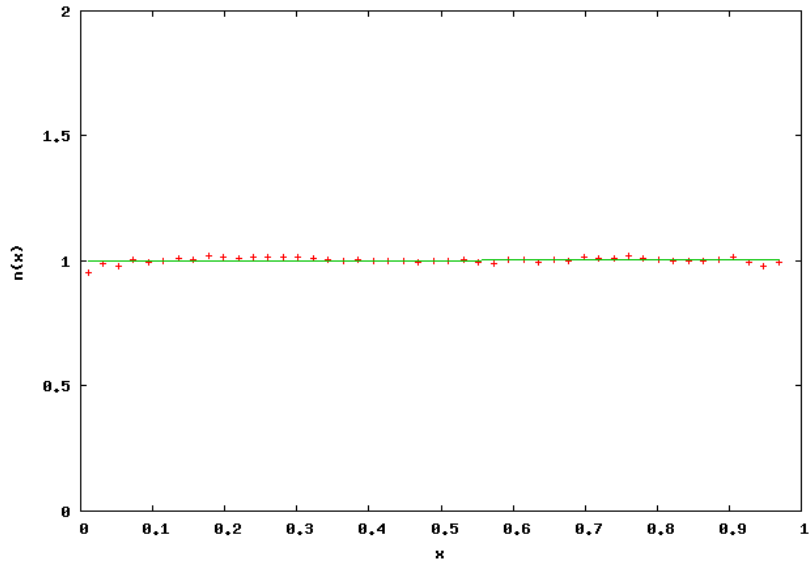

**Fig. 3.** The $n(x)$ relation (Poisson distribution; $N = 4000$).

## 2.4. Limits of Applicability of the Methods

We used the obtained artificial uniform distributions of points to study the algorithms applied to estimate the statistical quantities $\Gamma(r)$, $\Gamma*(r)$, $\eta_{ab}(r)$.

For uniform distributions, the conditional density and cylinder methods demonstrate good agreement of their analytical estimates of the parameters and of the method uncertainties with the results of numerical experiments.

Results from the conditional density method for scale lengths comparable to the distance to the nearest galaxies fluctuate strongly. The results from the cylinder method agreed with $\Gamma(r)$, $\Gamma*(r)$.

The cylinder method is fairly time-consuming, since the number of operations grows as $N^3$. It requires further study and development, being one way to overcome scale-length limitations following from the geometry of program samples.

# 3. CONSTRUCTING SAMPLES OF SDSS GALAXIES FOR ANALYSIS OF LARGE-SCALE STRUCTURE USING CORRELATION TECHNIQUES

## 3.1. Description and Statistics of the SDSS DR7 Data

When compiling galaxy surveys, a particular cosmological model is adopted. The galaxy's angular coordinates are then supplemented with its distance in the adopted model determined from its estimated recession velocity. However, measured velocities are not due solely to the Hubble expansion. They are distorted by the galaxy's peculiar velocity along the line of sight due to various gravitational effects. Therefore, the galaxy distances contain errors. Galaxy surveys based on velocities are called maps in "redshift space". Despite the fact that they represent a somewhat distorted version of the true 3D galaxy distribution, the distance errors are not serious enough to distort the overall pattern of the LSS.

The Sloan Digital Sky Survey was commenced about 10 years ago. The aims of the survey were to obtain CCD frames covering about 10 000 square degrees of the sky in five color filters and to perform spectroscopy of one million galaxies and 100 000 quasars in the same field. These aims were achieved with the seventh release of the SDSS (DR7). The images cover 11 663 square degrees in fields at low Galactic latitudes, or 2000 square degrees more than in the DR6. DR7 contains images of 357 million individual objects and 1.6 million measured spectra. These include 929 thousand spectra of galaxies, 121 thousand spectra of quasars, and 460 thousand spectra of stars [11]. The median redshift of the catalog is z ~ 0.07; most galaxies have z < 0.3 (the redshift distribution is shown in Fig. 4). The apparent magnitude range, corrected for extinction in the Galaxy in the $J$ filter, is $10.5 < m_J < 17.75$ (Fig. 5).

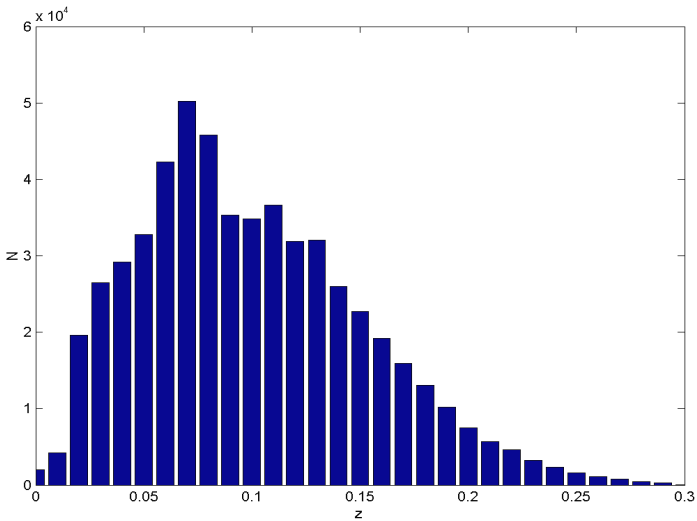

**Fig. 4.** Redshift distribution of the SDSS survey galaxies.

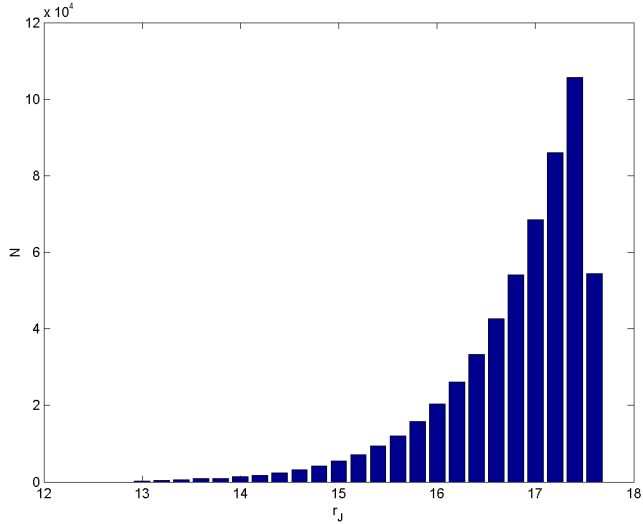

**Fig. 5.** Distribution of the SDSS survey galaxies over the apparent magnitude $r_J$.

### 3.2. Algorithm used to Construct a Volume-Limited Subsample

### 3.3.1. Boundaries of the region of the celestial sphere and redshift range.

We studied the LSS in the galaxy distribution using data from the SDSS DR7. We chose two regions from the survey: the largest closed area in the northern part of the celestial sphere, covering about 10 000 square degrees, and a narrow strip in the equatorial region (Fig. 6).

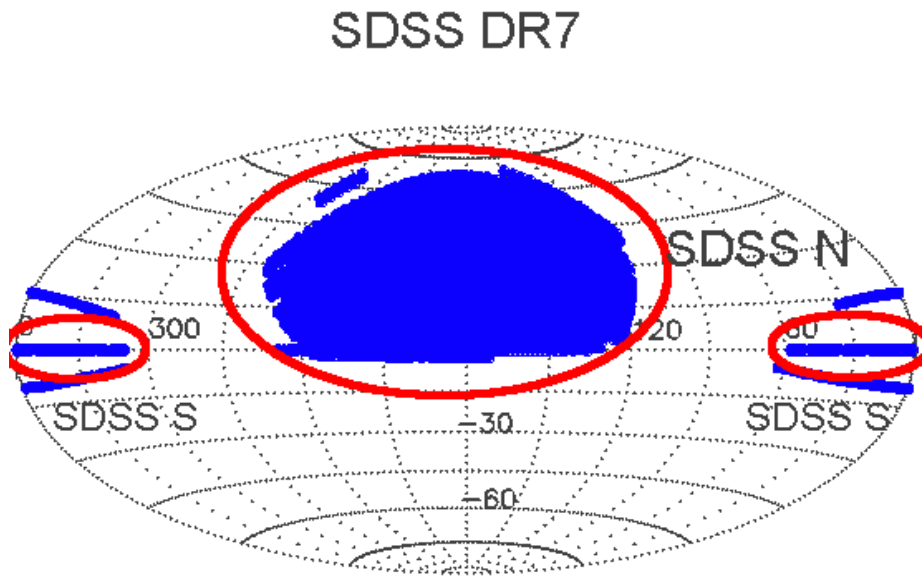

**Fig. 6.** Map of the northern and southern regions of the SDSS survey used

In addition to the limits on the coordinates on the celestial sphere, we adopted the following limits for the program galaxies: $z > 0.001$, to exclude nearby galaxies (because of their peculiar velocities), and $z < 0.3$, due to the survey depth. The number of galaxies in our sample is about 550000.

### 3.3.2. Determination of the metric distances to the galaxies and their absolute magnitudes.

To find the distances to the galaxies, we adopted a standard cosmological model with $\Omega_M = 0.3$ $\Omega_\Lambda = 0.7$. The Hubble constant was taken to be $H_0 = 100$ km s$^{-1}$ Mpc$^{-1}$. n the standard model, the metric distance is expressed as

$$d(z) = \frac{1}{H_0} \int_{\frac{1}{1+z}}^{1} \frac{dy}{y\left(\frac{\Omega_M}{y} + \Omega_\Lambda \cdot y^2\right)^{1/2}} \tag{22}$$

The derived distances correspond to so-called redshift space. In addition to the cosmological expansion velocity, the redshifts $z$ also include the peculiar velocities of galaxies and galaxy groups. The intrinsic galaxy distances correspond to intrinsic-distance space, and can be determined only using redshift independent methods.

We calculated the absolute magnitudes from the formula

$$M_r = m_r - 5\log_{10}[d(z)(1+z)] - k_r(z) - 25 \tag{23}$$

To find the $K$ correction, we used the formula [12]

$$k_r = z \cdot (2.61 - (m_g - m_r) - 0.64) \tag{24}$$

### 3.2.3. The volume-limited subsamples.

To take into account selection effects due to the magnitude limit of the SDSS DR7 galaxies, $m < 17.7$, we chose a distance $l_{max}$ and the corresponding maximum absolute magnitude observed at this distance, $M_{max}$. In this case, the sample will contain only those galaxies ($l < l_{max}$, $M < M_{max}$) that are guaranteed to be visible in the whole volume of the subsample.

In each of the regions, we selected five subsamples bounded by absolute magnitudes in different ranges ($M_{min}$; $M_{max}$). For each subsample in the equatorial region, we adopted only $M_{max}$. $z - M$ diagrams for these chosen regions are presented in Figs. 7 and 8. The parameters of the samples are collected in Table 1.

**Table 1.** Parameters of the volume-limited samples.

| Name | N | $\alpha_{min}$ | $\alpha_{max}$ | $\delta_{min}$ | $\delta_{max}$ | $d_{min}$ Mpc | $d_{max}$ Mpc | $R_{max}$ Mpc | $M_{min}$ | $M_{max}$ | $\overline{Distance}$ Mpc |
|---|---|---|---|---|---|---|---|---|---|---|---|
| N1 | 12140 | 132 | 234 | 0 | 57 | 9.2 | 131.2 | 41.6 | -18.0 | -19.0 | 1.58 |
| N2 | 11081 | 170 | 220 | 10 | 50 | 17.7 | 246.0 | 60.6 | -19.5 | -20.0 | 2.26 |
| N3 | 12593 | 180 | 220 | 10 | 50 | 26.0 | 370.0 | 90.0 | -20.5 | -21.0 | 2.97 |
| N4 | 11346 | 180 | 220 | 10 | 50 | 70.0 | 560.6 | 135.7 | -21.5 | -22.0 | 4.96 |
| N5 | 13435 | 140 | 220 | 0 | 57 | 62.4 | 671.7 | 212.1 | -22.0 | -22.5 | 8.56 |
| S1 | 1260 | -45 | 58 | -1.0. | 1.0 | 12.9 | 125.7 | 2.3 | -18 | - | 1.15 |
| S2 | 2914 | -45 | 58 | -1.0 | 1.0 | 19.5 | 196.2 | 3.6 | -19 | - | 1.28 |
| S3 | 3795 | -45 | 58 | -1.0 | 1.0 | 30.2 | 298.8 | 5.2 | -20 | - | 1.96 |
| S4 | 3298 | -45 | 58 | -1.0 | 1.0 | 61.2 | 448.9 | 7.95 | -21 | - | 3.21 |
| S5 | 863 | -45 | 58 | -1.0 | 1.0 | 113.0 | 665.0 | 11.7 | -22 | - | 9.41 |

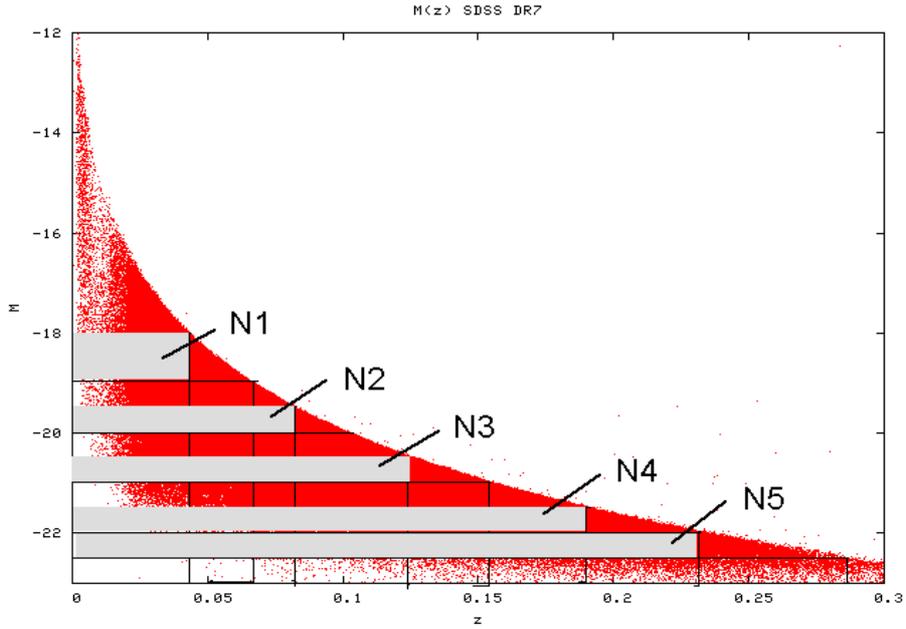

**Fig. 7.** *M—z* diagram for the SDSS DR7 northern sample.

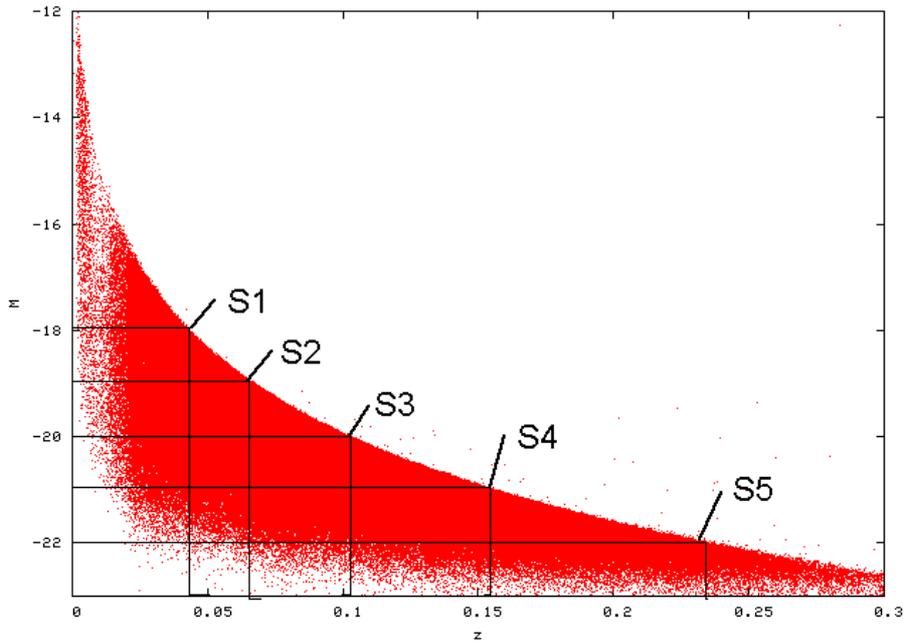

**Fig. 8.** *M—z* diagram for the SDSS DR7 southern sample.

## 4. CORRELATION ANALYSIS RESULTS FOR THE SDSS DR7 DATA

### 4.1. Conditional Density

Conditional density methods enable analysis of the scale length $r \sim 100$ Mpc (for the N5 sample). Plots of the conditional density $\Gamma^*(r)$ and the conditional density in shells $\Gamma(r)$ are presented in Figs. 9–12. Table 2 contains the results for the $\Gamma^*(r)$ and $\Gamma(r)$ methods; the $[r_{min}, r_{max}]$, range used to obtain the estimate is indicated for each sample. For all the samples, $\Gamma^*(r)$ demonstrates an obvious slope change at scale lengths exceeding 10–15 Mpc, where it enters a flat interval. This behavior could mean that the galaxy distribution becomes uniform at such scale lengths [13] or that large-radius spheres are located in a small volume of space where some structure is present, i.e., a systematic deviation is encountered [14]. In this case, the density decrease will be slower. We observe a strong deviation in the behavior of the estimated parameters for the S5 sample compared to the other samples. This is most likely due to presence of a small number of galaxies in a relatively large volume, which makes the fluctuations of the derived parameters fairly large.

**Table 2.** $D$ values estimated using the conditional density method

| Sample | $r_{min}(Mpc)$ | $r_{max}(Mpc)$ | $D, \Gamma*(r)$ | $D, \Gamma(r)$ |
|---|---|---|---|---|
| N1 | 0.5 | 40.1 | 2.175 | 2.223 |
| N2 | 0.5 | 58.1 | 2.137 | 2.175 |
| N3 | 0.5 | 85.4 | 2.037 | 2.074 |
| N4 | 0.5 | 128.6 | 2.115 | 2.029 |
| N5 | 0.5 | 199.5 | 2.035 | 2.039 |
| S1 | 0.1 | 2.3 | 2.124 | 1.731 |
| S2 | 0.1 | 3.5 | 2.237 | 1.970 |
| S3 | 0.1 | 5.3 | 1.998 | 1.867 |
| S4 | 0.1 | 7.7 | 2.040 | 1.726 |
| S5 | 0.1 | 11.4 | 1.782 | 1.000 |

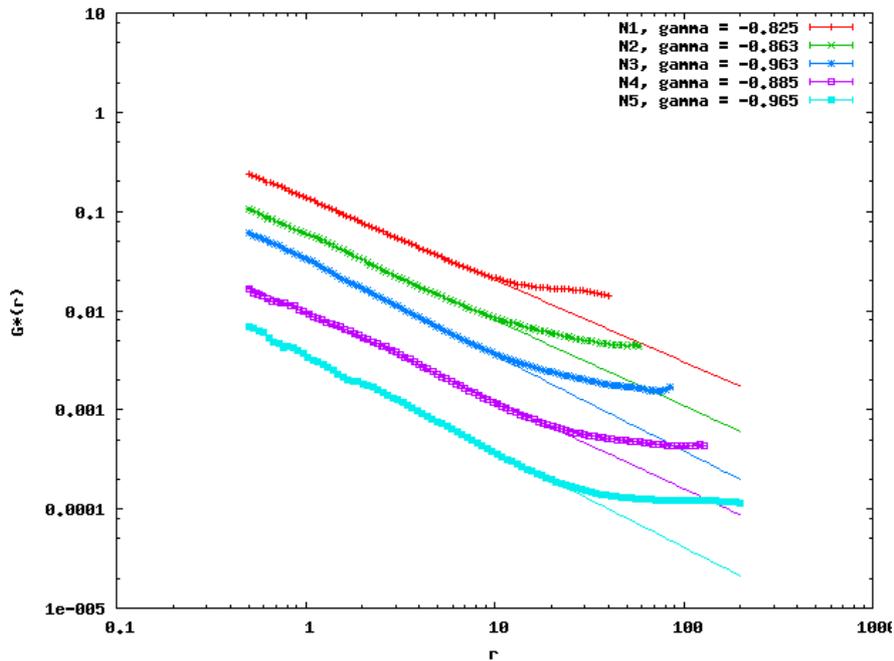

**Fig. 9.** $\Gamma*(r)$ relations for the complete SDSS northern samples.

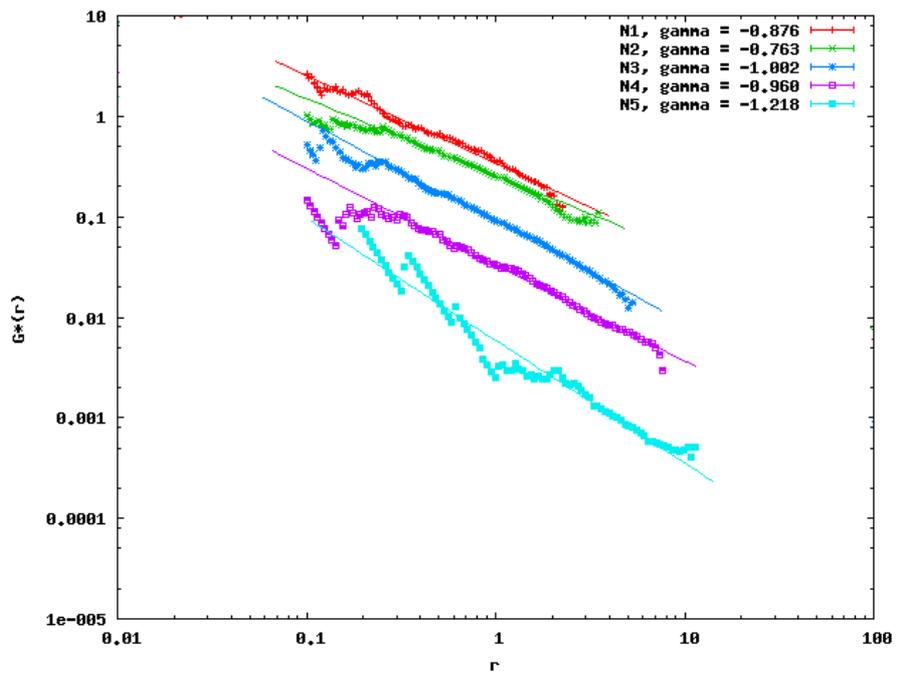

**Fig. 10.** $\Gamma^*(r)$ relations for the complete SDSS southern samples.

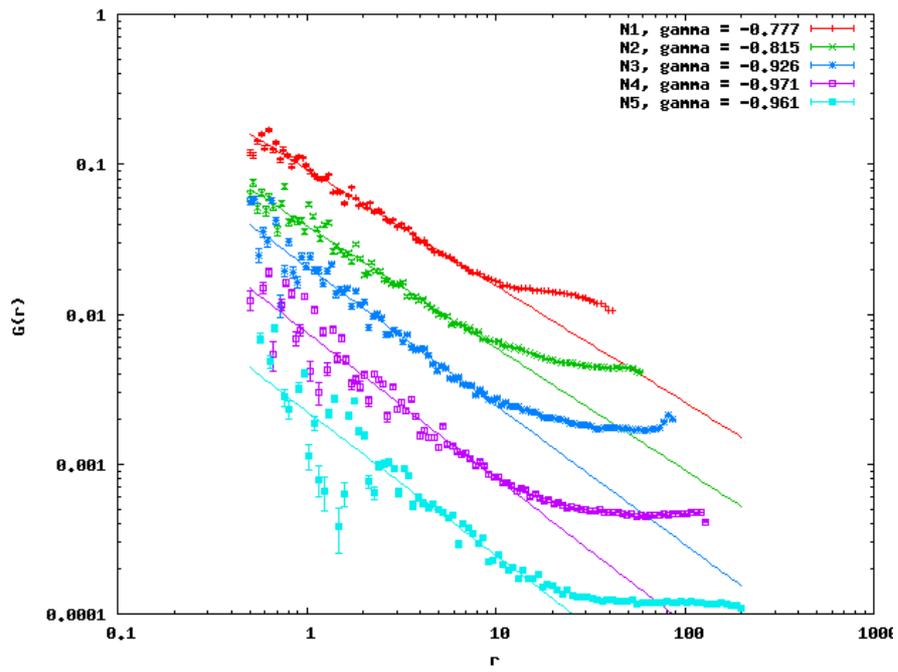

**Fig. 11.** $\Gamma(r)$ relations for the complete SDSS northern samples

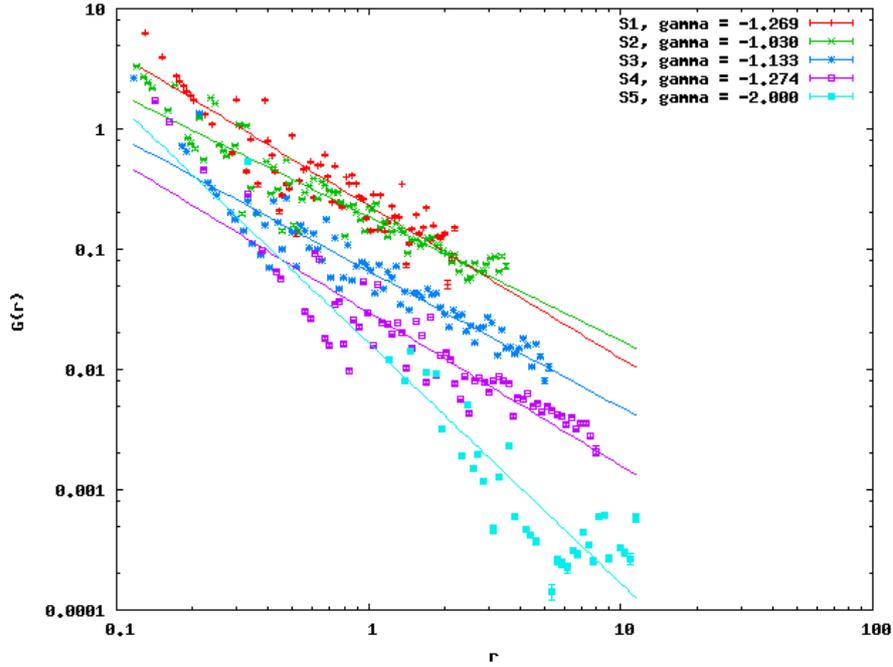

**Fig. 12.** Γ (*r*) relations for the complete SDSS southern samples.

### 4.2. The Cylinder Method

We obtained results with the cylinder method for each sample adopting three radius values for the cylinder: 0.25, 0.5, 1.5 Mpc. The results were computed for several ranges of cylinder lengths, each consisting of cylinders with lengths varying within 10% of a given value. The diagrams for the samples, lengths, and radii are shown in Figs. 13 and 14.

As an example, Table 3 presents the results obtained using (18) and (19) for the N1 and S1 samples. For scale lengths *r* < 70 Mpc, the results are in good agreement with Γ(*r*) and Γ$^*$ (*r*).

Since this method is mathematically exact, the theoretically best results should correspond to an infinitely small cylinder radius. Thus, increasing the radius contaminates the true pattern of the galaxy distribution along the axis. Consequently, we obtain a fractal dimension that is too large; i.e., a distribution that is more uniform.

The results for the northern and southern samples generally agree. The slight deviations could be due to the different absolute-magnitude limitations for the complete samples. These were adopted because the number of galaxies in the southern SDSS region is much smaller than that in the northern region. Thus, the southern statistics in the northern magnitude intervals would have been insufficient. As in the conditional density method, the results can be considerably distorted for the brightest galaxies of the southern region (the S5 sample), since the number of galaxies in this sample is small.

One explanation for such distortions on large scale lengths could be that the direction distribution does not stay isotropic for long cylinders: they only fit the sample volume in a particular direction. Thus there are not enough cylinders in other directions, we obtain no real averaging of the data. If a structure exists in a certain direction, all cylinders of a given length will cover it, and we will obtain the galaxy distribution only in this structure, rather than in the whole sample after averaging. This flaw of the cylinder method can be overcome by using surveys with a larger effective depth and a wider coverage of the celestial sphere.

Figures 13 and 14 display examples of the distribution of *N* (*x*) /*N* for the N1 and N3, S1 and S3

samples, for cylinder radii of 0.25 and 1 Mpc and lengths up to 30 and 90 Mpc.

Table 3. Main parameters of the cylinder method derived for the N1 and S1 samples

| Sample | Radius Mpc | Length Mpc | D | R1 | R2 | Sample | Radius Mpc | Length Mpc | D | R1 | R2 |
|---|---|---|---|---|---|---|---|---|---|---|---|
| N1 | 0.25 | 30 | 2.201 | 481.481 | 0 | S1 | 0.25 | 30 | 2.295 | 564.392 | 0 |
| | | 50 | 1.951 | 316.100 | 0.010 | | | 50 | 2.013 | 302.401 | 0.002 |
| | | 70 | 1.750 | 280.767 | 0.020 | | | 70 | 1.865 | 223.449 | 0.004 |
| | | 90 | 1.575 | 264.347 | 0.025 | | | 90 | 1.460 | 191.095 | 0.019 |
| | | 110 | 1.406 | 251.298 | 0.028 | | | 110 | 1.285 | 182.868 | 0.023 |
| | | 130 | 1.362 | 268.966 | 0.031 | | | 130 | 1.292 | 177.336 | 0.021 |
| | | 150 | 1.446 | 291.568 | 0.030 | | | 150 | 1.429 | 201.806 | 0.022 |
| | 0.5 | 30 | 2.396 | 976.867 | 0 | | 0.5 | 30 | 2.499 | 1440.853 | 0 |
| | | 50 | 2.130 | 473.910 | 0.003 | | | 50 | 2.244 | 548.155 | 0 |
| | | 70 | 1.900 | 368.857 | 0.018 | | | 70 | 2.048 | 307.502 | 0 |
| | | 90 | 1.720 | 333.604 | 0.025 | | | 90 | 1.695 | 242.667 | 0.018 |
| | | 110 | 1.547 | 307.983 | 0.029 | | | 110 | 1.445 | 215.521 | 0.024 |
| | | 130 | 1.467 | 312.801 | 0.031 | | | 130 | 1.396 | 195.503 | 0.022 |
| | | 150 | 1.515 | 333.621 | 0.031 | | | 150 | 1.556 | 218.175 | 0.021 |
| | 1.0 | 30 | 2.649 | 5869.228 | 0 | | 1.0 | 30 | 2.705 | 11271.946 | 0 |
| | | 50 | 2.360 | 1106.735 | 0 | | | 50 | 2.413 | 1292.129 | 0 |
| | | 70 | 2.112 | 642.160 | 0.014 | | | 70 | 2.182 | 459.813 | 0 |
| | | 90 | 1.998 | 516.310 | 0.024 | | | 90 | 1.896 | 332.782 | 0.016 |
| | | 110 | 1.732 | 439.217 | 0.030 | | | 110 | 1.753 | 312.303 | 0.022 |
| | | 130 | 1.642 | 426.559 | 0.032 | | | 130 | 1.569 | 238.919 | 0.023 |
| | | 150 | 1.668 | 447.890 | 0.032 | | | 150 | 1.737 | 273.333 | 0.020 |

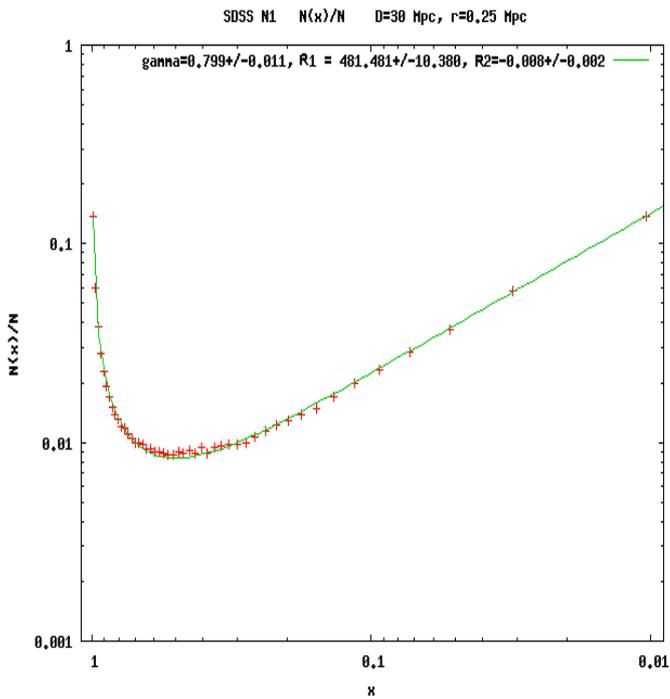
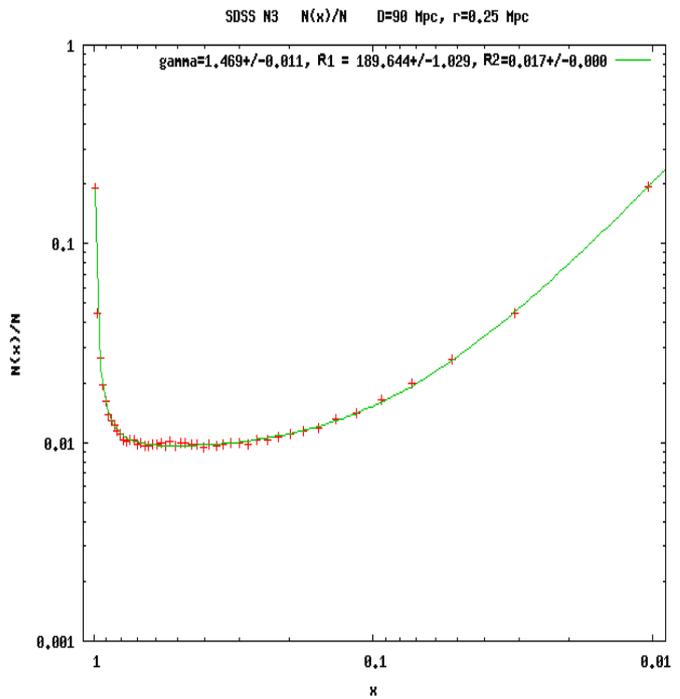

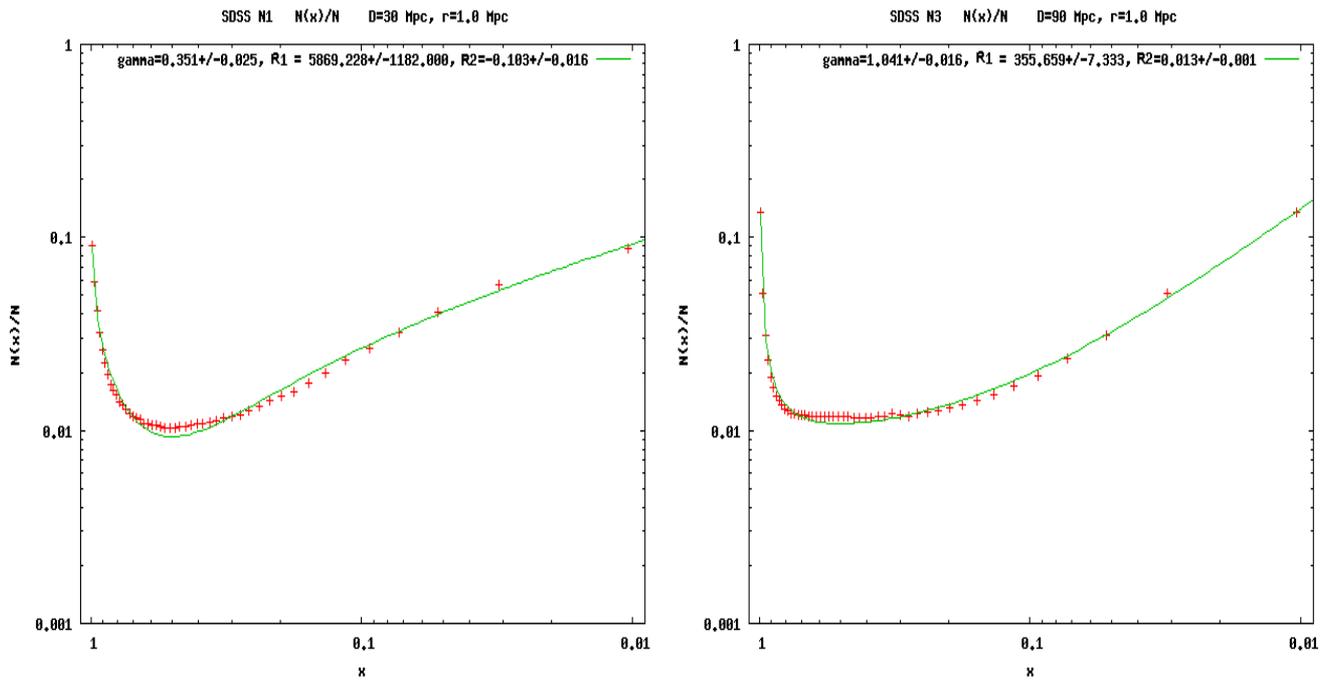

Fig. 13. Distribution of N (x) /N for the N1 and N3 samples with radii of 0.25 and 1Mpc and lengths up to 30 and 90Mpc.

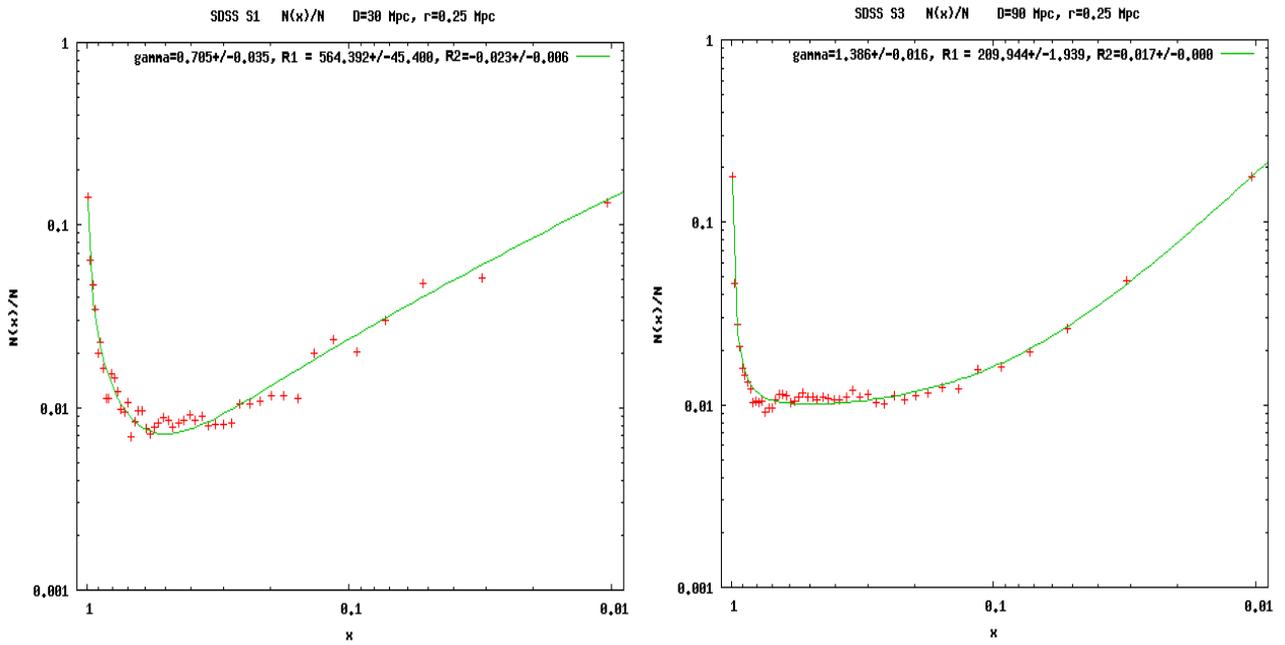

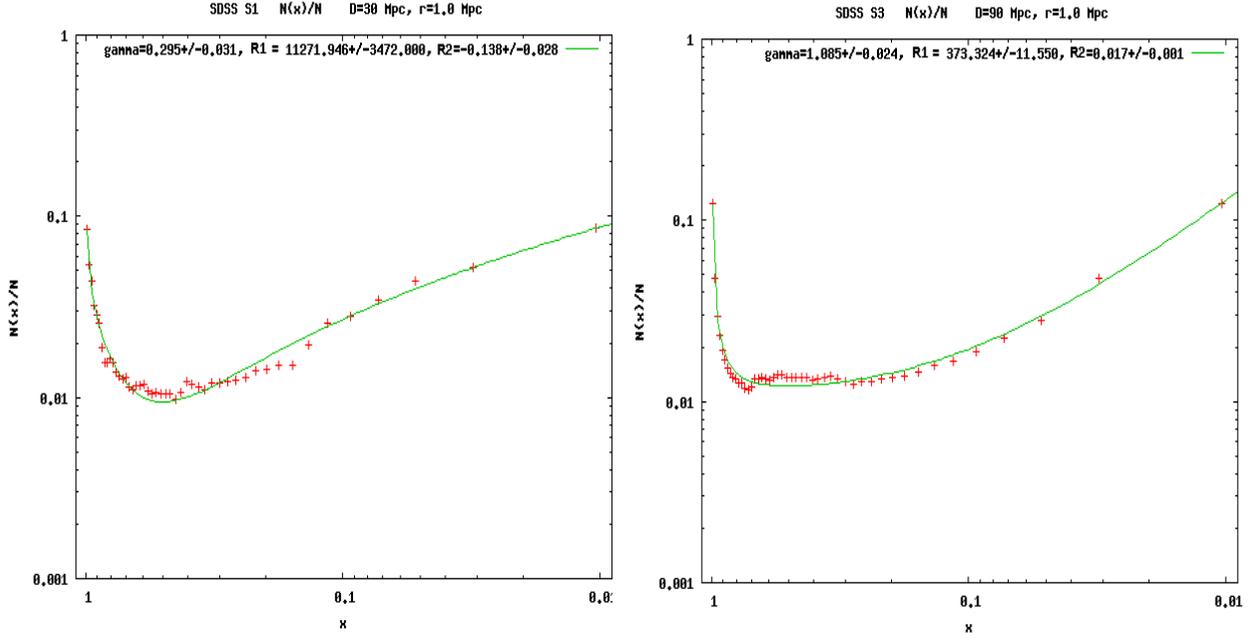

Fig. 14. Distribution of N (x) /N for the S1 and S3 samples with radii of 0.25 and 1Mpc and lengths up to 30 and 90Mpc.

### 4.3. Radial Distributions

To check the results obtained with the above methods and look for structures capable of influencing the results, we plotted the radial distributions of the number $N(d)$ and density $n(d)$ versus distance for each sample. Thus, we can find deviations of the observed density from its theoretical estimate and draw conclusions about the presence of a density peak or dip, i.e., on the presence of structures or voids.

To improve the resolution of the method, we subdivided each of the northern SDSS subsamples into eight sectors in ($\alpha, \delta$), and the narrow southern strip into four sectors. Each sector was subdivided into volumes, and we derived distributions of the number and density of galaxies in each of the volumes as functions of the distance to the volume. In this case, the number of galaxies and the volume are differential quantities, which increase as the square of the distance. Thus, the distributions of the number and density of galaxies can be theoretically approximated with a quadratic relation (and with a constant in the case of a uniform distribution). The deviation of the observed galaxy density from the theoretical approximation is represented

$$(\sigma_{obs})^2 = (\sigma_p)^2 + (\sigma_{corr})^2 \qquad (25)$$

where $\sigma_p$ are Poisson fluctuations and $\sigma_{corr}$ are fluctuations related to the presence of structures. Thus, the amplitude of the deviations related to structures can be found from the formula

$$(\sigma_{corr})^2 = (\sigma_{obs})^2 - (\sigma_p)^2 \qquad (26)$$

The Poisson deviations of the number and density of galaxies are given by

$$\Delta N_{th} = \pm \sigma_P * N = \frac{N}{\sqrt{(N)}} = \sqrt{(N)} \qquad (27)$$

$$\Delta n_{th} = \pm \frac{1}{\sqrt{(N)}} \frac{1}{V} \cdot N = \frac{\sqrt{(N)}}{V} \tag{28}$$

If $\Delta N_{obs} = |N_{obs} - N_{th}|$ or $\Delta n_{obs} = |n_{obs} - n_{th}|$ is larger or smaller than $\Delta N_{th}$ or $\Delta n_{th}$ the density of galaxies is higher or lower than that for the uniform distribution. The amount of inhomogeneity can be estimated from the number of data points in the diagram above or below the theoretical value.

We found that each of the sectors in each of the SDSS northern subsamples displayed a density excess above the uniform level at a scale length corresponding to the linear part of the density variation. The size of these fluctuations is 20–50 Mpc for scale lengths below 100 Mpc, 50–100 Mpc for scale lengths from 100 to 300 Mpc, and 100–200 Mpc for scale lengths exceeding 300 Mpc. Thus, the size of the structures increases with the scale length, leading us to reject the hypothesis that homogeneity is the reason the conditional density diagram enters its flat interval.

The S5 sample on scale lengths of 0–10 Mpc corresponding to sphere radii that fit into this sample reveals voids with densities appreciably lower than the uniform level. This explains why the conditional density method finds a lower fractal dimension for this than for the other samples. As an example, Fig. 15 shows distributions of the number and density of galaxies for the S5 sample. The numerical data characterizing the largest inhomogeneities in each of the sectors are presented in Table 4.

**Fig. 15.** Distributions of *N*(*d*) and *n*(*d*) for the S5 sample. The regions above the curve correspond to detected structures and those under the curve to voids. The curves *1* are $N_{obs}$ and $n_{obs}$; the curves *2* are $N_{th}$ and $n_{th}$; and curves *3* are $N_{th}$ and $n_{th}$ with their Poisson uncertainties.

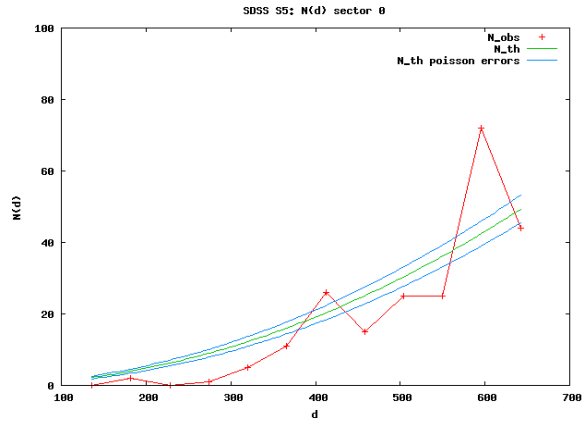
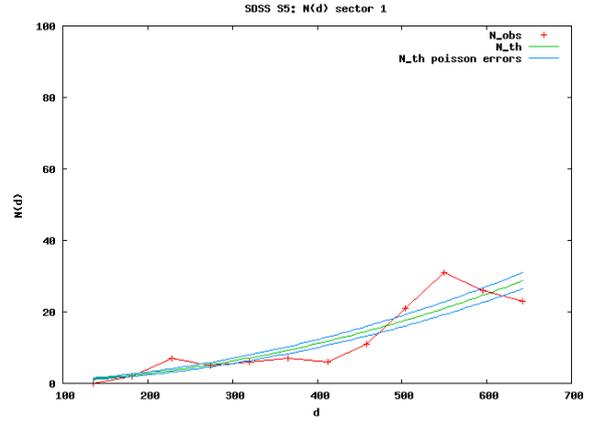
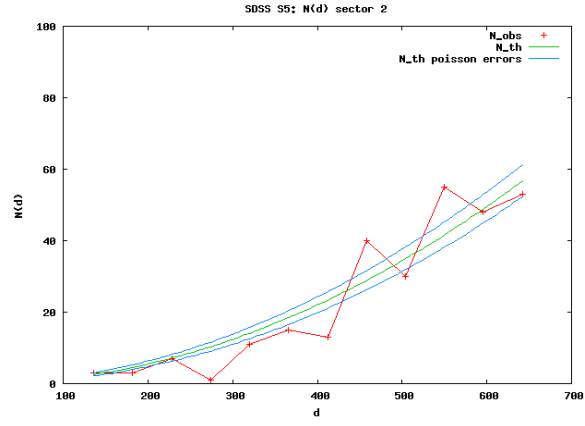
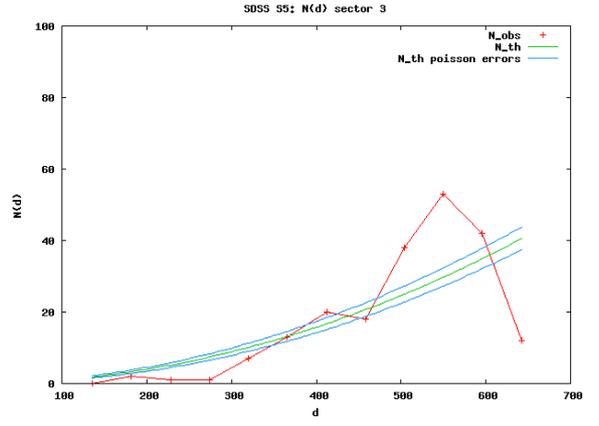
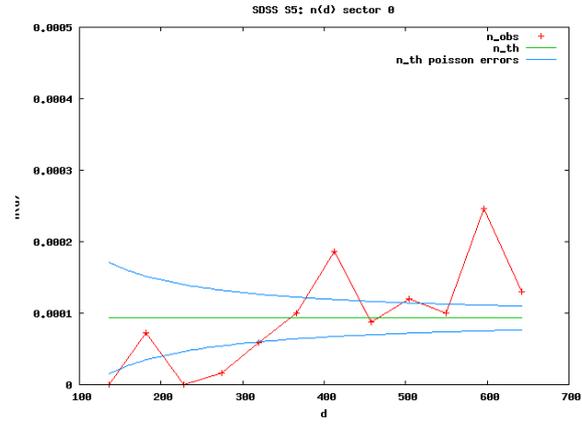
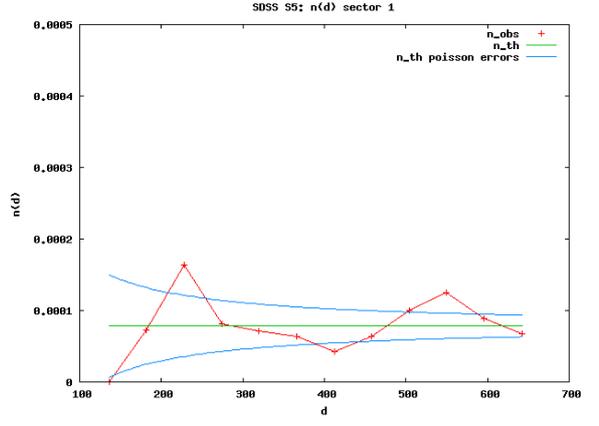
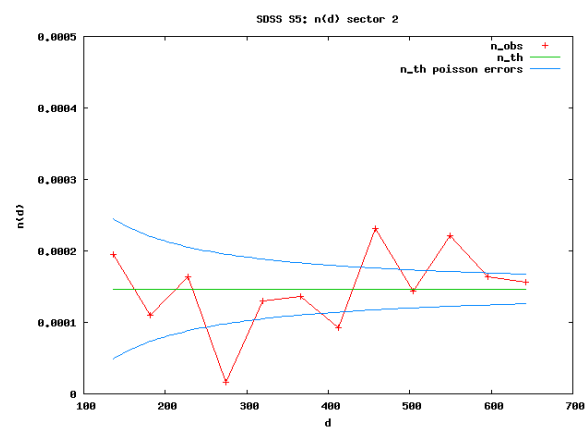
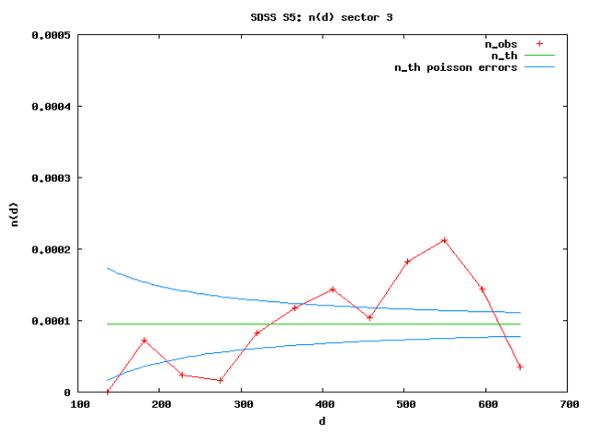

**Table 4.** Maximum density $n_{max}$ (in $\sigma_P$), size (in Mpc), and distance to inhomogeneities (in Mpc) derived using the radial distribution method.

| Sample | Par | sec0 | sec1 | sec2 | sec3 | sec4 | sec5 | sec6 | sec7 |
|---|---|---|---|---|---|---|---|---|---|
| N1 | n_max | 6 | 5 | 6 | 7 | 4 | 5 | 3 | 5 |
|  | size | 30 | 20 | 20 | 50 | 10 | 10 | 20 | 85 |
|  | dist | 85 | 65 | 70 | 100 | 80 | 105 | 75 | 20 |
| N2 | n_max | 5 | 5 | 3 | 5 | 4 | 3 | 5 | 3 |
|  | size | 25 | 30 | 30 | 50 | 20 | 25 | 40 | 30 |
|  | dist | 60 | 60 | 60 | 195 | 100 | 60 | 175 | 200 |
| N3 | n_max | 5 | 5 | 5 | 2 | 3 | 4 | 10 | 5 |
|  | size | 75 | 100 | 100 | 25 | 50 | 25 | 50 | 50 |
|  | dist | 200 | 240 | 180 | 160 | 175 | 210 | 176 | 200 |
| N4 | n_max | 4 | 4 | 2 | 3 | 4 | 10 | 8 | 4 |
|  | size | 50 | 50 | 25 | 75 | 50 | 50 | 75 | 100 |
|  | dist | 490 | 500 | 200 | 450 | 500 | 500 | 490 | 500 |
| N5 | n_max | 6 | 4 | 5 | 3 | 3 | 3 | 5 | 4 |
|  | size | 150 | 100 | 50 | 100 | 50 | 50 | 100 | 150 |
|  | dist | 550 | 500 | 650 | 450 | 380 | 480 | 550 | 500 |
| S1 | n_max | 5 | 2 | 5 | 4 |  |  |  |  |
|  | size | 20 | 10 | 10 | 10 |  |  |  |  |
|  | dist | 80 | 60 | 50 | 100 |  |  |  |  |
| S2 | n_max | 5 | 5 | 5 | 4 |  |  |  |  |
|  | size | 20 | 20 | 20 | 20 |  |  |  |  |
|  | dist | 145 | 175 | 130 | 120 |  |  |  |  |
| S3 | n_max | 5 | 4 | 7 | 4 |  |  |  |  |
|  | size | 25 | 25 | 30 | 25 |  |  |  |  |
|  | dist | 150 | 175 | 130 | 100 |  |  |  |  |
| S4 | n_max | 7 | 2 | 2 | 5 |  |  |  |  |
|  | size | 50 | 25 | 25 | 50 |  |  |  |  |
|  | dist | 400 | 175 | 175 | 400 |  |  |  |  |
| S5 | n_max | 4 | 2 | 2 | 4 |  |  |  |  |
|  | size | 100 | 100 | 50 | 100 |  |  |  |  |
|  | dist | 600 | 550 | 450 | 550 |  |  |  |  |

## 5. CONCLUSIONS

We have analyzed the spatial distribution of galaxies from the latest release of the Sloan Digital Sky Survey of galaxy redshifts (DR7 SDSS), using the complete correlation function (conditional density), two-point conditional density (cylinder), and radial density methods. Our analysis demonstrated that the conditional density has a power-law form on scale lengths 0.5–30 Mpc/h, with the power-law index corresponding to the fractal dimension $D = 2.2 \pm 0.2$.

This quantity display an essentially flat interval on scale lengths exceeding 30 Mpc/h, interpreted in [13] as resulting from a uniform spatial distribution of galaxies. However, it was demonstrated in [15] using artificial galaxy catalogs with different fractal dimensions that a flat interval in the conditional density also appeared in the case of a purely fractal distribution, as an artifact of the finite sample volume. A new statistical method for the analysis of the spatial distribution of galaxies suggested recently (SL statistics) [1,14] can be used to detect the violation of translational invariance of the galaxy distribution to scale lengths of 300 Mpc/h, with the sample depth of the SDSS galaxies being 600 Mpc/h. Thus, the presence of a flat interval in the conditional density curve can probably be explained by the presence of giant inhomogeneous structures that partially enter the volume of the analyzed galaxy sample. This is supported by our analyses using the cylinder method and radial count method. The cylinder method indicates that the power-law form of the conditional density continues to scale lengths of 70 Mpc/$h$ with $D = 2.0 \pm 0.3$. For large scale lengths, the presence of un averaged structures in the galaxy distribution begins to distort the estimated fractal dimension. The radial density method indicates inhomogeneities in the spatial distribution of galaxies with a scale length of 200 Mpc/$h$ and a density contrast of two, confirming the recently established violation of statistical homogeneity in deep samples of SDSS galaxies [1,14].